\documentclass{article}

\usepackage{cmap} 
\usepackage{krctran}

\newcounter{fpage}
\usepackage[ruled,linesnumbered]{algorithm2e}
\usepackage{float}

\usepackage[dvipsnames]{xcolor} 

\graphicspath{ {./pictures/} }

\usepackage{arydshln} 

\usepackage{mathtools} 

\usepackage[hidelinks]{hyperref}
\usepackage{bookmark}

\usepackage[shortcuts]{extdash} 

\usepackage[figurename=Fig.]{caption}
\usepackage[tablename=Table]{caption}
\usepackage[labelsep=period]{caption}
\usepackage[labelfont=small,labelfont=it]{caption}
\usepackage[textfont=small]{caption}

\EngVer 

\SetKwRepeat{Do}{do}{while} 
\SetKwComment{Comment}{}{}

\SetCommentSty{mycommfont}

\theoremstyle{definition}
\newtheorem{definition}{Definition}[section]

\newif\ifmonochrome

\monochromefalse 

\ifmonochrome
    \newcommand{\vgreen}{\color{black}}
    \newcommand{\vviolet}{\color{black}}
    \newcommand{\vred}{\color{black}}
\else
    \newcommand{\vgreen}{\color{OliveGreen}}
    \newcommand{\vviolet}{\color{violet}}
    \newcommand{\vred}{\color{red}}
\fi

\newcommand{\vI}{ \ensuremath{\vgreen v_1}}
\newcommand{\vbI}{ \ensuremath{\vgreen \boldsymbol{ v_1}}}

\newcommand{\vIII}{ \ensuremath{\vgreen v_3}}
\newcommand{\vbIII}{ \ensuremath{\vgreen \boldsymbol{ v_3}}}

\newcommand{\vIV}{ \ensuremath{\vgreen v_4}}
\newcommand{\vbIV}{ \ensuremath{\vgreen \boldsymbol{ v_4}}}

\newcommand{\vbV}{ \ensuremath{\vgreen \boldsymbol{ v_5}}}

\newcommand{\vIIsynI}{ \ensuremath{ v^2_{\vred syn_1}}}

\newcommand{\vbIsynI}{ \ensuremath{ \boldsymbol{ v^1_{\vred syn_1}}}}

\newcommand{\vIsynII}{ \ensuremath{ v^1_{\vviolet syn_2}}}
\newcommand{\vbIsynII}{ \ensuremath{ \boldsymbol{ v^1_{\vviolet syn_2}}}}

\newcommand{\vIIsynII}{ \ensuremath{ v^2_{\vviolet syn_2}}}
\newcommand{\vbIIsynII}{ \ensuremath{ \boldsymbol{ v^2_{\vviolet syn_2}}}}

\newcommand{\CIepsilon}{ \ensuremath{ C_{\vred 1}(\epsilon) }}

\newcommand{\CIIepsilon}{ \ensuremath{ C_{\vviolet 2}(\epsilon) }}

\newcommand*\mean[1]{\overline{#1}}

\begin{document}

\procname{
Transactions of Karelian Research Centre   RAS  \ \ \ \ \ \ \ \ \  Труды Карельского научного центра РАН
\\  No. 7. 2018.  P.~149--163 \hspace{4.6cm}      \textnumero~7. 2018. С.~{149}--{163}\\ {\bf DOI: 10.17076/mat829}}
\udk{УДК 81.32}

\rustitle{Алгоритм решения WSD-задачи на основе \protect\\ 
нового способа вычисления близости \protect\\ контекстов с~учетом $\varepsilon$-фильтрации слов}

\engtitle{WSD algorithm based on a new method of \protect\\ vector-word contexts proximity calculation \protect\\ via $\varepsilon$-filtration}

\rusauthor{А.~Н.~Кириллов, Н.~Б.~Крижановская, А.~А.~Крижановский}
\engauthor{A.~N.~Kirillov, N.~B.~Krizhanovskaya, A.~A.~Krizhanovsky}

\organization{Institute of Applied Mathematical Research of the Karelian Research Centre \protect\\ of the Russian Academy of Sciences}

\rusabstract{
Рассмотрена задача разрешения лексической многозначности (WSD), а именно: по известным наборам синонимов (синсеты) и предложений с этими синонимами требуется автоматически определить, в каком значении использовано слово в предложении. 
Экспертами были размечены 1285 предложений, выбрано одно из заранее известных значений (синсетов). 
Для решения WSD-задачи предложен алгоритм, основанный на новом способе вычисления близости контекстов. При этом для более высокой точности выполняется предварительная $\varepsilon$-фильтрация слов, как в предложении, так и в наборе синонимов. 
%
Проведена обширная программа экспериментов. Реализовано четыре алгоритма, включая предложенный. Эксперименты показали, что в ряде случаев новый алгоритм показывает лучшие результаты.
Разработанное программное обеспечение и размеченный корпус с открытой лицензией доступны онлайн. Использованы синсеты Викисловаря и тексты Викитеки.
Краткое описание работы в виде слайдов доступно по ссылке (https://goo.gl/9ak6Gt), 
видео с докладом также доступно онлайн (https://youtu.be/-DLmRkepf58).
}

\engabstract{
The problem of word sense disambiguation (WSD) is considered in the article. 
Set of synonyms (synsets) and sentences with these synonyms are taken. It is necessary to automatically select the meaning of the word in the sentence.
1285 sentences were tagged by experts, namely, one of the dictionary meanings was selected by experts for target words.
To solve the WSD problem, an algorithm based on a new method of vector-word contexts proximity calculation is proposed. 
A preliminary $\varepsilon$-filtering of words is performed, both in the sentence and in the set of synonyms, in order to achieve higher accuracy. 
An extensive program of experiments was carried out. 
Four algorithms are implemented, including the new algorithm. 
Experiments have shown that in some cases the new algorithm produces better results. 
The developed software and the tagged corpus have an open license and are available online. Wiktionary and Wikisource are used.
A brief description of this work can be viewed as slides (https://goo.gl/9ak6Gt).
A video lecture in Russian about this research is available online (https://youtu.be/-DLmRkepf58).
}

\ruskeywords{синоним; синсет; корпусная лингвистика; word2vec; Викитека; WSD; RusVectores; Викисловарь.}
\engkeywords{synonym; synset; corpus linguistics; word2vec; Wikisource; WSD; RusVectores; Wiktionary.}

\maketitle
\begin{articletext}

%

\section{Introduction}

The problem of word sense disambiguation (WSD) is a real challenge to computer scientists and linguists.
Lexical ambiguity is widespread and is one of the obstructions in natural language processing.

In our previous work ``Calculated attributes of synonym sets''~\cite{krizhanovsky2018calculated}, we have proposed the geometric approach to mathematical modeling of synonym set (synset) using the word vector representation. 
Several geometric characteristics of the synset words were suggested (synset interior, synset word rank and centrality). They are used to select the most significant synset words, i.e. the words whose senses are the nearest to the sense of the synset.

%
The topic related to polysemy, synonyms, filtering and WSD is continued in this article. 
Let us formulate the mathematical foundations for solving the problems of computational linguistics in this article.

Using the approach proposed in the paper~\cite{Chen2014unified}, we present the WSD algorithm based on a new context distance (proximity) calculation via $\varepsilon$\=/filtration.
The experiments show the advantages of the proposed distance over the traditional average vectors similarity measure of distance between contexts.

\section{New $\varepsilon$-proximity between finite sets} \label{Section:NewEpsProximity}

It is quite evident that the context distance choice is one of the crucial factors influencing WSD algorithms.
Here, in order to classify discrete structures, namely contexts, we propose a new approach to  context proximity based on
Hausdorff metric and symmetric difference of sets:  $A\triangle B = (A\cup B)\setminus (A \cap B)$.

\begin{figure}[H]
   \centering
\ifmonochrome
\else
    \includegraphics[keepaspectratio=true,width=0.6\columnwidth]{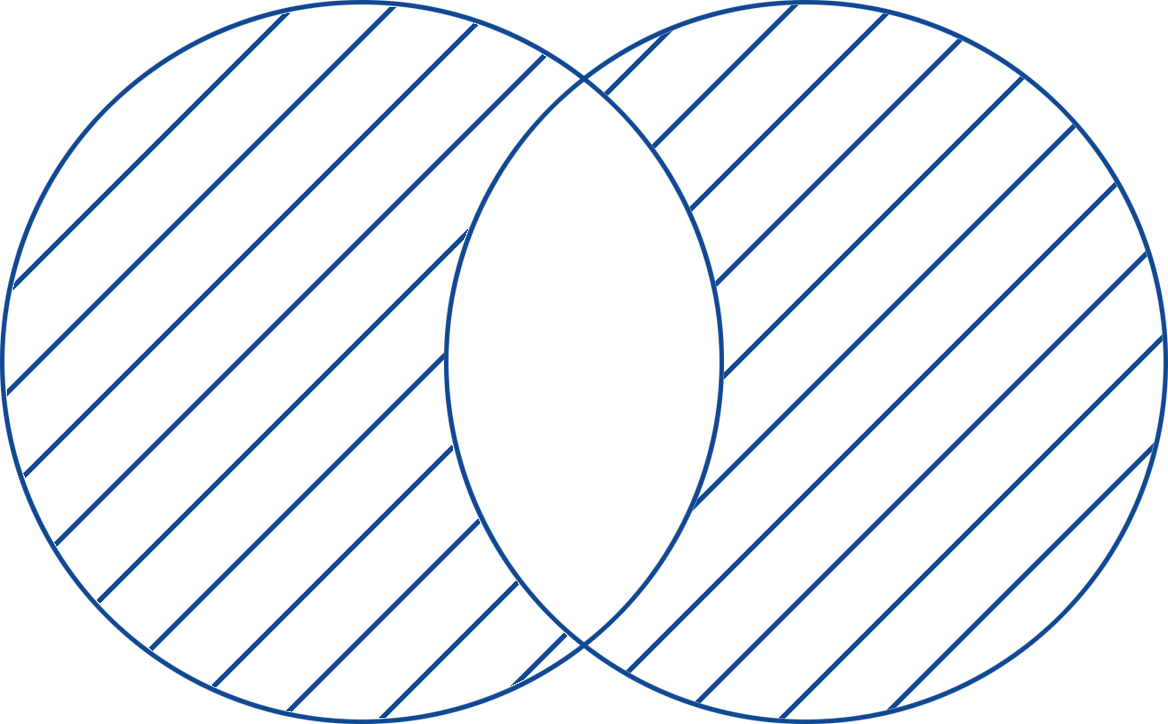}
\fi
    \caption{The set $A\triangle B$ is the shaded part of circles}
    \label{fig:OutOfIntersection}
\end{figure}

Recall the notion of Hausdorff metric.
Consider a metric space $(X, \varrho)$ where $X$ is a set, $\varrho$ is a metric in $X$.
Define the $\varepsilon$-dilatation $A+\varepsilon$ of a set  $A\subset X$

$$
A+\varepsilon=\cup\{\mean{B}_\varepsilon(x): x\in A\},
$$
where $\mean{B}_\varepsilon(x)$ is a closed ball centered at $x$ with the radius $\varepsilon$.

The Hausdorff distance
$\varrho_H(A, B)$ between compact nonempty sets $A$ and $B$ is
$$
\varrho_H(A, B)=\min\{\varepsilon > 0:  (A \subset B+\varepsilon)\wedge (B\subset A+\varepsilon)\},
$$
where  $A+\varepsilon$, $B+\varepsilon$ are the $\varepsilon$-dilatations of $A$ and $B$.
Consider the following sets~(Fig.~\ref{fig:EpsDilatation}):
$$
A(\varepsilon)=A\cap (B+\varepsilon), \quad B(\varepsilon)=B\cap (A+\varepsilon).
$$

\begin{figure}[H]
   \centering
\ifmonochrome
\else
    \includegraphics[keepaspectratio=true,width=0.8\columnwidth]{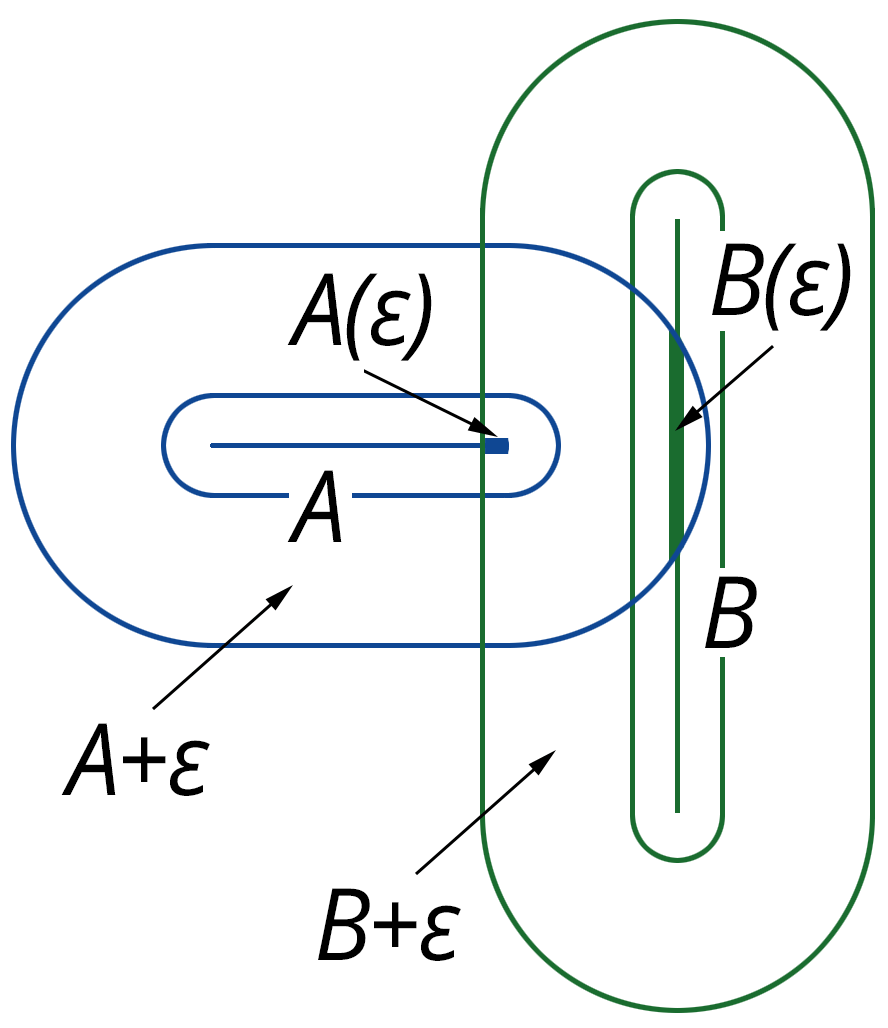}
\fi
    \caption{Two sets $A+\varepsilon$ and $B+\varepsilon$ are the $\varepsilon$\=/dilatations of segments $A$ and $B$, and two new proposed set-valued maps $A(\varepsilon)$ and $B(\varepsilon)$ were inspired by Hausdorff distance}
    \label{fig:EpsDilatation}
\end{figure}

Then 
$$
\varrho_H(A, B)=\min\{\varepsilon > 0: A(\varepsilon)\cup B(\varepsilon)=A\cup B\}.
$$
Consider two contexts $W_1=\{w_{11},...,w_{1m}\}$, $W_2=\{w_{21},...,w_{2n}\}$, where $w_{1i}, \ w_{2j}$ are words in the contexts,  $i=1,..,m, \ j=1,...,n$.
Denote by $V_1=\{v_{11},...,v_{1m}\}$, $V_2=\{v_{21},...,v_{2n}\}$ the sets of vectors $v_{1i}, \ v_{2j}$
corresponding to the words $w_{1i}, \ w_{2j}$.
Recall that generally in WSD procedures, the distance between words is measured by similarity function, 
which is a cosine of angle between vectors representing words: ${sim(v_1, v_2)= \frac{(v_1, v_2)}{||v_1||||v_2||}}$,
where  $(v_1, v_2)$ is a scalar (inner) product of vectors $v_1, v_2$, and $||v_i||$\,is a norm of vector, ${i=1, 2}$. 
In what follows,  ${ sim(v_1, v_2) \in [-1, 1]}$.
Thus, the less distance the more similarity. Keeping in mind the latter remark, we introduce the following $\varepsilon$\=/proximity of vector contexts $V_1, \ V_2$.
Given ${\varepsilon\geq 0}$, construct the sets
$$
C(V_1, V_2, \varepsilon)=\{u, v:  u\in V_1, v\in V_2, sim(u, v) \geq \varepsilon\}.
$$
$$
D(V_1, V_2, \varepsilon)=(V_1\cup V_2)\setminus C(V_1, V_2).
$$
Supposing that $sim$ plays the role of a metric, then $C(V_1, V_2, \varepsilon)$ is analogous to the expression ${A(\varepsilon)\cup B(\varepsilon)}$ 
in the definition of the Hausdorff distance.

Denote by $|Y|$ the power of a set $Y\subset X$,  $\mathbb{R}_+=\{x: x\geq 0, x \in \mathbb{R}\}$.

\theoremstyle{definition}
\begin{definition}{}
The $K$-proximity of contexts $V_1, V_2$ is the function 
$$
K(V_1, V_2, \varepsilon)= \frac{|C(V_1, V_2, \varepsilon)|}{|V_1\cup V_2|}.
$$
\end{definition}
It is clear that  $K(V_1, V_2, \varepsilon) \in [0, 1]$.
We also define the following function.

\begin{definition}{}
The $\tilde{K}$-proximity of contexts $V_1, V_2$ is the function 
$$
\tilde{K}(V_1, V_2, \varepsilon)=\frac{|C(V_1, V_2, \varepsilon)|}{1+|D(V_1, V_2, \varepsilon)|},
$$
describing the ratio of ``near'' and ``distant'' elements of sets. 
\end{definition}
The definition implies that 
${\min \tilde{K}(V_1, V_2, \varepsilon)=0}$,  
${\max \tilde{K}(V_1, V_2, \varepsilon)=|V_1\cup V_2|}$. 
The presence of 1 in the denominator permits to avoid zero denominator when $|D(V_1, V_2, \varepsilon)|=0$.

The ubiquitous distance $\varrho$ between contexts $V_1, V_2$ is based on the similarity of average vectors: $\varrho(V_1, V_2)=sim(\mean{V}_1, \mean{V}_2)$.
But the example (Fig.~\ref{fig:AverageEvil}) shows that for two geometrically distant and not too similar structures ${\varrho(V_1, V_2)=1}$, that is the similarity $\varrho$ takes the maximum value.

\subsection{Example}

Consider the sets $A = \{a_1, a_2, a_3\}$, $B = \{b_1\}$ pictured in Fig.~\ref{fig:AverageEvil}, 
where $a_1+a_3=\overrightarrow{0}$,  $a_2=b_1$. 
Then, 
$sim (A, B) = sim (\frac{1}{3}(a_1+a_2+a_3), b_1)=sim (a_2, b_1)=1$,
$\tilde{K}(A, B, \varepsilon)=\frac{2}{3}$, $K(A, B, \varepsilon)=\frac{1}{2}$.

The equality of average vectors does not mean 
the coincidence of $A$ and $B$, 
which are rather different~(Fig.~\ref{fig:AverageEvil}).

\begin{figure}[H]
   \centering
    \includegraphics[keepaspectratio=true,width=0.95\columnwidth]{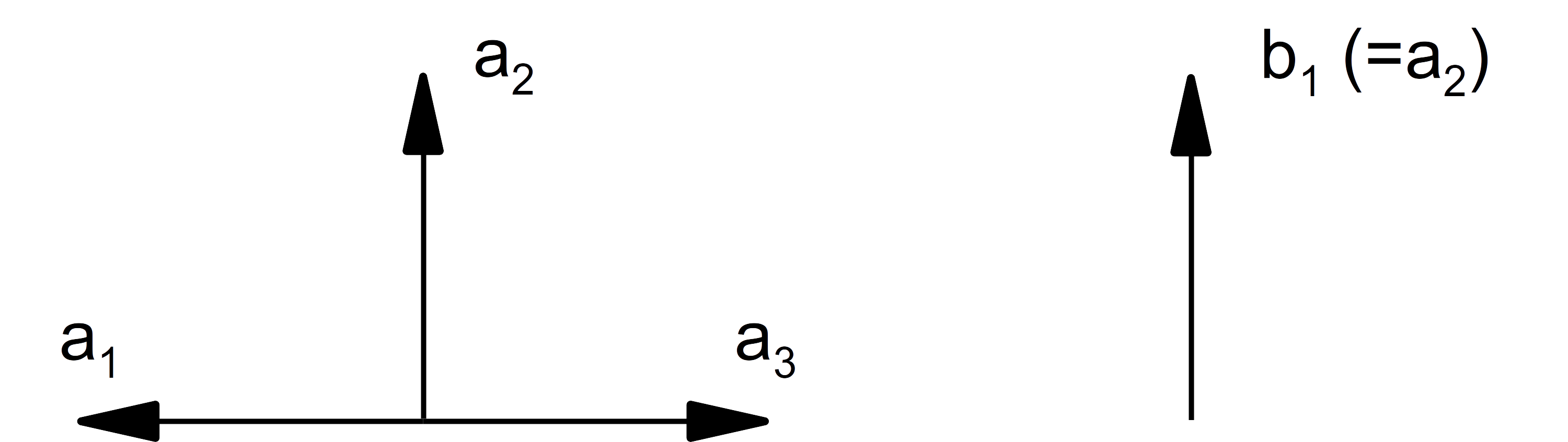}
    \caption{An example of similar average vectors 
    ($\mean{A} = a_2 = b_1 = \mean{B}$) and totally different sets of vectors: $\{a_1, a_2, a_3\}$ and $\{b_1\}$}
    \label{fig:AverageEvil}
\end{figure}


\section{Average algorithm with synonyms $\varepsilon$\=/filtration} \label{Section:FirstAlgorithm}

Consider a sentence $S_w=(w_1 \dots w_i^* \dots w_n)$ 
containing a target word $w_i^*$ (denote it as $w^*$).
and a vector representation
$S=(v_1 \dots v_i^* \dots v_n)$ of $S_w$, 
where $w_j$ is a word, $v_j$ is a vector representation of $w_j$. 
Denote $v_i^*$ as $v^*$.
Suppose the target  word $w^*$ has $l$  senses. 
Denote by $syn_k^w$ a synset corresponding to $k$-th sense,
$k=1, \dots ,l$,  $syn_k^w=\{w_{k1},\dots,w_{ki_k}\}$, where $w_{kp}$ are synonyms. 
Let $syn_k=\{v_{k1},\dots,v_{ki_k}\}$ be a set of vector representations of synonyms  $w_{kp}$,  $p=1,\dots,i_k.$

In what follows, we introduce a procedure of $\varepsilon$\=/filtration, the idea of which is borrowed from the paper~\cite{Chen2014unified}.

The synset filtration is the formation of a so called candidate set which consists of those synonyms whose similarity with the words from a sentence is higher than a similarity threshold\,$\varepsilon$.

The first average algorithm~\ref{alg:first}, described below, uses average vectors of words of sentences and average vectors of the candidate set of synonyms in synsets.

This algorithm contains the following lines.

Line \ref{alg1line:aver_sentence}. Calculate the average vector of words of the sentence $S$ 
$$
\mean{S}=\frac{1}{n}\sum_{j=1}^nv_j
$$

Lines \ref{alg1line:epsilon}--\ref{alg1line:set_size}. Given $\varepsilon>0$, let us construct the filtered set of synonyms for each synset
$$
cand_k(\varepsilon)=\{u\in syn_k: u \neq v^*, \  sim(u, v^*)>\varepsilon\}.
$$
Denote by $s_k(\varepsilon)=|(cand_k(\varepsilon))|$ the power of a set $cand_k(\varepsilon)$.

Line \ref{alg1line:average_synset}. Calculate for $s_k(\varepsilon)>0$ the average vector of the synset candidates
$$
\mean{syn}_k(\varepsilon)=\frac{1}{s_k(\varepsilon)}\sum_{u\in cand_k(\varepsilon)}u.
$$
If $s_k(\varepsilon)=0$, then let $\mean{syn}_k(\varepsilon)$ be equal to the zero vector.

Line \ref{alg1line:similarity}. Calculate the similarity of the average vectors of the sentence and the \textit{k}-th filtered synset
$$
sim_k(\varepsilon)=sim(\mean{syn}_k(\varepsilon), \mean{S}).
$$

Line \ref{alg1line:sim_max}--\ref{alg1line:unique}. Suppose $max_{k=1,\dots,l}\{sim_k(\varepsilon)\}=sim_{\scalebox{0.9}{$k^*$}}(\varepsilon)$, i.e. $k^*\in \{1,\dots,l\}$ is the number of the
largest  $sim_k(\varepsilon)$. If $k^*$ is not unique, then take another $\varepsilon >0$ and repeat the procedure from line~\ref{alg1line:epsilon}.

Result: the target word $w^*$ has the sense corresponding to the $k^*$-th synset $syn_{k^*}^w$.

\begin{algorithm*}
    \caption{Average algorithm with synonyms $\varepsilon$\=/filtration}
    \label{alg:first}
\DontPrintSemicolon
\SetAlgoLined
    \KwData{$v^*$ -- vector of the target word $w^*$ with $l$ senses (synsets), \hfill \break
        $v_i \in S$, $S$ -- sentence with the target word $w^*$, $v^* \in S$, \hfill \break
        $\{syn_k\}$ -- synsets of the target word, that is $syn_k \ni v^*$, $k = \overline{1,l}$.
       }
\KwResult{$k^*\in \{1,\dots,l\}$ is the number of the sense of the word $w^*$ in the sentence $S$.}
\BlankLine
    $ \mean{S}=\frac{1}{n}\sum\limits_{j=1}^{n} {v_j}$\label{alg1line:aver_sentence}, \Comment{the average vector of words of the sentence $S$}
    \Do{$k^*$ is not unique}{
        take $\varepsilon>0$ \label{alg1line:epsilon}\;
        \Comment{foreach synset of the target word}
        \ForEach{$syn_k \ni v^*$}{%
            \Comment{construct the filtered set $cand_k(\varepsilon)$ of the synset $syn_k$:}
            $cand_k(\varepsilon)=\{u\in syn_k: u \neq v^*, \  sim(u, v^*)>\varepsilon\}$\;
            $s_k(\varepsilon) = |cand_k(\varepsilon)|$, \label{alg1line:set_size} \Comment{number of candidates of synonyms}
            \Comment{the average vector of synset candidates:}
            $\mean{syn}_k(\varepsilon)= 
                \begin{cases}
                    \frac{1}{s_k(\varepsilon)}\sum\limits_{u\in cand_k(\varepsilon)}u,& \text{\small if $s_k(\varepsilon)>0$} \\
                    \overrightarrow{0},& \text{\small if $s_k(\varepsilon)=0$}
                \end{cases}$ \label{alg1line:average_synset} \;
            \Comment{the similarity of average vectors of the sentence and the \textit{k}-th filtered synset:}
            $sim_k(\varepsilon)=sim(\mean{syn}_k(\varepsilon), \mean{S})$ \label{alg1line:similarity} \;
        }
        $sim_{\scalebox{0.9}{$k^*$}}(\varepsilon) = \max_{k=1,\dots,l} \{sim_k(\varepsilon)\} \Rightarrow k^*\in \{1,\dots,l\}$ \label{alg1line:sim_max}, \Comment{$k^*$ is the number of the largest $sim_k(\varepsilon)$}
    } \label{alg1line:unique}
\end{algorithm*}

Remark: in the case $\epsilon=0$, we denote this algorithm as $\mean{A}_0$-algorithm. In this case, the traditional averaging of similarity is used. 

\textit{Note}. $\mean{A}_0$-algorithm was used in our experiments, 
it was implemented in Python.\footnote{
See the function \textit{selectSynsetForSentenceByAverageSimilarity} in the file \url{https://github.com/componavt/wcorpus.py/blob/master/src/test\_synset\_for\_sentence/lib\_sfors/synset\_selector.py}}

\subsection{$\mean{A}_0$-algorithm example}

A simple example and figures~\ref{fig:WiktSlushit}--\ref{fig:CirclesSentSynsetSkeleton} will help to understand how this $\mean{A}_0$-algorithm works. 

Take some dictionary word $w_2$ with several senses and several synonym sets 
(for example, $syn_1$ and $syn_2$) and the sentence $S$ with this word (Fig.~\ref{fig:WiktSlushit}). The task is to select a meaning (synset) of $w_2$ (that is the target word is $w_2^*$) used in the sentence $S$ via the $\mean{A}_0$\=/algorithm. 

Let us match the input data and the symbols used in the $\mean{A}_0$-algorithm. The word ``служить'' (sluzhit') corresponds to the vector $v_2$. 


\begin{figure}[H]
   \centering
\ifmonochrome
\else
    \includegraphics[keepaspectratio=true,width=0.99\columnwidth]{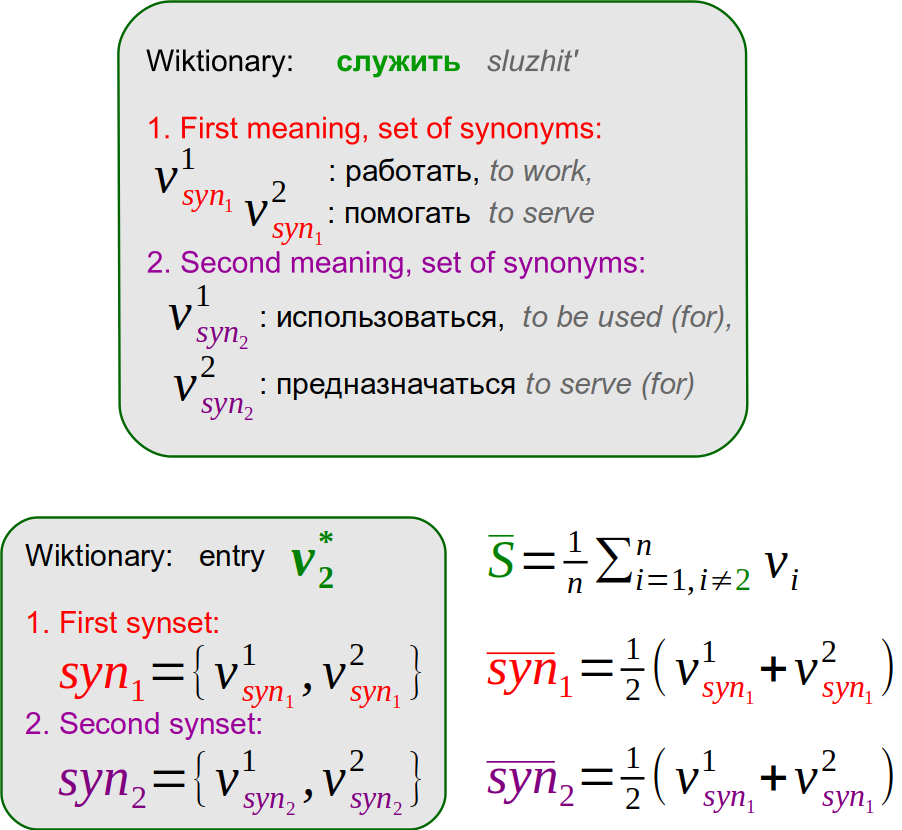}
\fi
    \caption{Digest of the Wiktionary entry ``служить'' (sluzhit') and mean vectors $\mean{syn_1}$ and $\mean{syn_2}$ of the synonyms sets $syn_1$, $syn_2$ and the sentence $S$ with this word $w_2^*$}
    \label{fig:WiktSlushit}
\end{figure}


\begin{figure}[H]
   \centering
\ifmonochrome
\else
    \includegraphics[keepaspectratio=true,width=0.99\columnwidth]{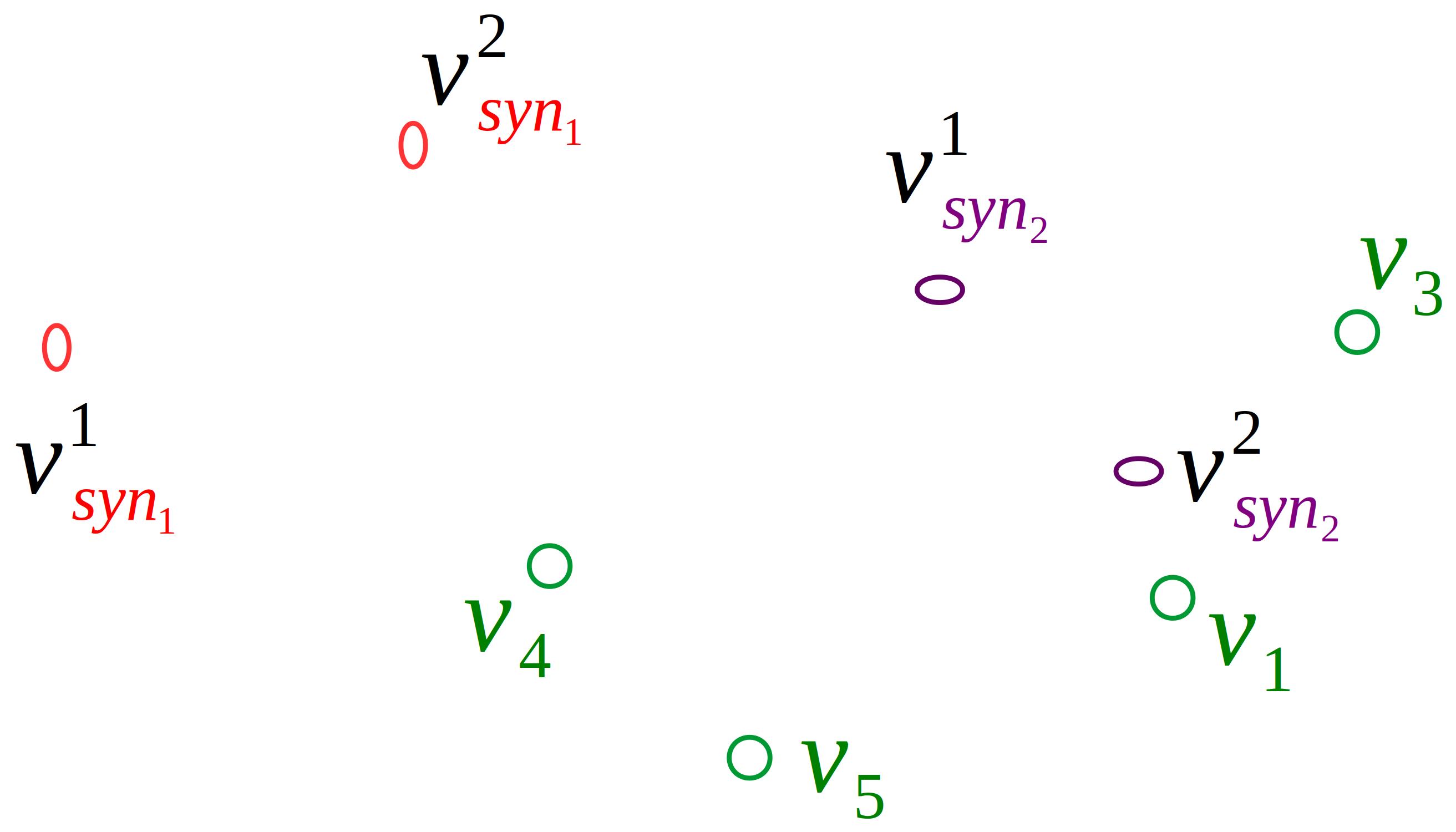} 
\fi
    \caption{Sample source data are 
        (1)~vertices $v_1...v_5$ corresponding to words of the sentence $S$,  
            the vertex $v_2$ was excluded since it corresponds to the target word $w_2^*$, 
        and (2)~the target word $w_2^*$ 
            with two synsets $syn_1$ and $syn_2$ (Fig.~\ref{fig:WiktSlushit}), 
        (3)~vertices (vectors correspond to words) of the first synset are 
            $\{v^1_{syn_1}, v^2_{syn_1}\}$ 
            and the second synset -- $\{v^1_{syn_2}, v^2_{syn_2}\}$}
    \label{fig:VerticesSource}
\end{figure}

\begin{figure}[H]
   \centering
\ifmonochrome
\else
    \includegraphics[keepaspectratio=true,width=0.99\columnwidth]{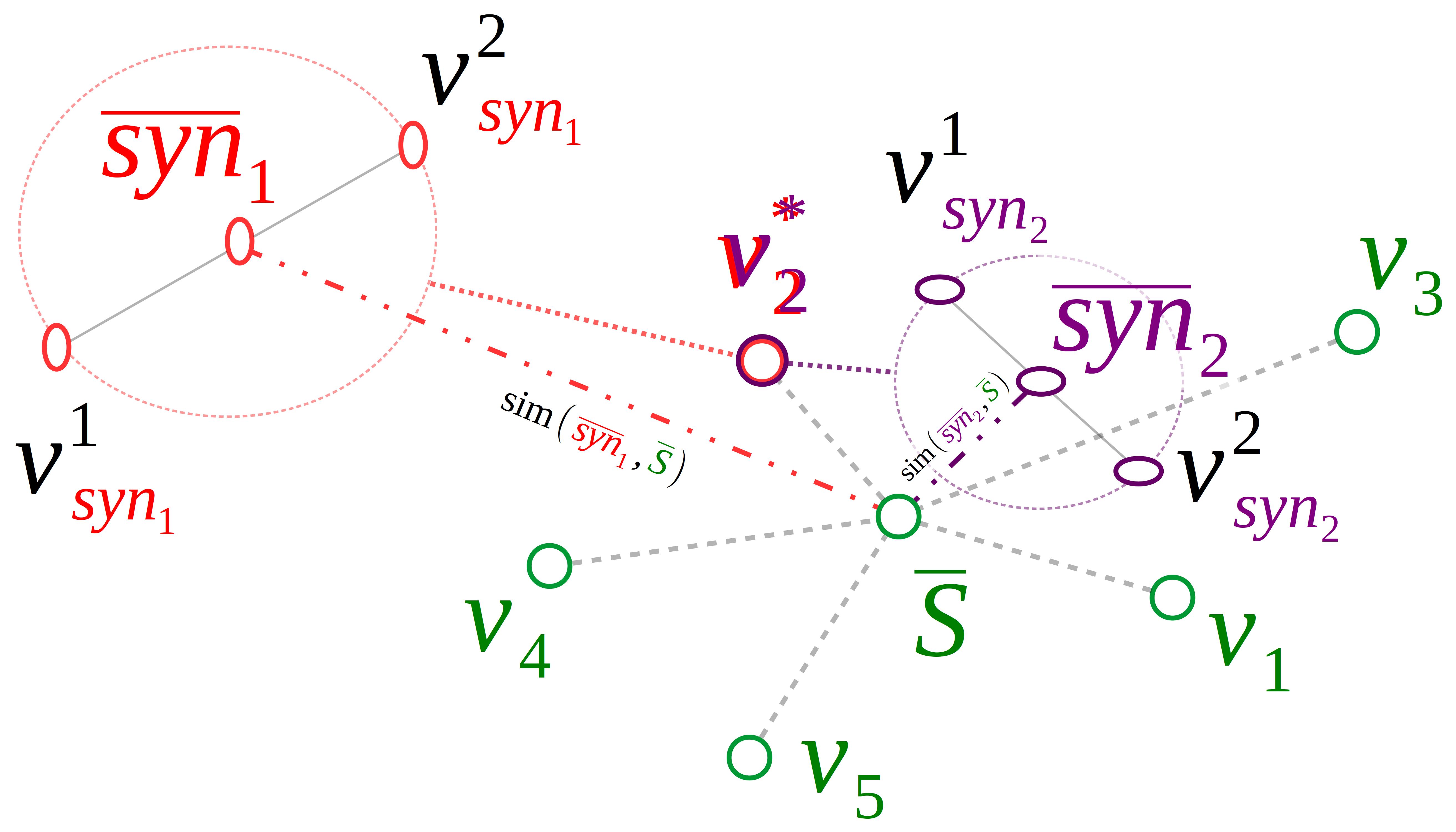} 
\fi
    \caption{Similarity between the mean value of vectors of the sentence 
    and the first synonym set is lower than the similarity with the second synset, 
    that is $ sim ( \mean{syn}_1, \mean{ S } ) < sim ( \mean{syn}_2, \mean{ S } ) $. 
    Thus, the second sense of the target word $w_2^*$ (the second synset $syn_2$) will be selected in the sentence $S$ by $\mean{A}_0$\=/algorithm}
    \label{fig:CirclesSentSynsetSkeleton}
\end{figure}

\bfullwidth
\efullwidth

There is a dictionary article about this word in the Wiktionary, 
see Fig.~\ref{fig:WiktSlushit} (a parsed database of Wiktionary is used in our projects).\footnote{See section ``Web of tools and resources'' on page~\pageref{Section:WebOfTools}.}

Two synonym sets of this Wiktionary entry are denoted by $syn_1$ and $syn_2$.

Mean values of the vectors corresponding to synonyms in these synsets 
will be denoted as ${\mean{ syn }}_1$ and $ {\mean{ syn }}_2$, 
and $ \mean{ S }$ is the mean vector of all vectors corresponding to words in the sentence $S$ containing the word ``служить'' (sluzhit').


\section{Average algorithm with sentence and synonyms $\varepsilon$-filtration ($\mean{A}_\varepsilon$)}

This algorithm~\ref{alg:aver_eps} is a modification of algorithm~\ref{alg:first}. 
The filtration of a sentence is added to synset filtration.
Namely, we select a word from the sentence for which the similarity 
with at least one synonym from the synset is higher than the similarity threshold $\varepsilon$. 
Then, we average the set of selected words forming the set of candidates from the sentence.
Let us explain algorithm~\ref{alg:aver_eps} line by line.

Lines \ref{alg2line:epsilon}--\ref{alg2line:sentence_size}. Given $\varepsilon >0$, 
let us construct the set of words of the sentence $S$ 
filtered by synonyms of the \textit{k}-th synset $syn_k$
\begin{multline*}
cand_kS(\varepsilon) = \{v \in S: \exists u\in syn_k, sim(v, u)>\varepsilon, \\
    \ v \neq v^*, u \neq v^*\}
\end{multline*}
Denote by  $S_k(\varepsilon)=|cand_kS(\varepsilon)|$ the power of the~set $cand_kS(\varepsilon)$.

Line \ref{alg2line:average_sentence}. Calculate the average vector of words of the filtered sentence
$$
\mean{{cand}_kS}(\varepsilon)=\frac{1}{S_k(\varepsilon)}\sum_{v\in cand_kS(\varepsilon)}v
$$
If $S_k(\varepsilon)=0$, then let $\mean{{cand}_kS}(\varepsilon)$ be equal to the zero vector.

Lines \ref{alg2line:synset_filtered}--\ref{alg2line:set_size}. Construct filtered sets of synonyms
\begin{multline*}
cand\ syn_k(\varepsilon)=\{u\in syn_k: \exists v\in S, sim(u, v)>\varepsilon, \\
    u \neq v^*, v \neq v^* \}.
\end{multline*}

Denote by $s_k(\varepsilon)=|cand\ syn_k(\varepsilon)|$ the power of the \textit{k}-th filtered synonym set.

Line \ref{alg2line:average_synset}. Calculate for $s_k(\varepsilon)>0$ the average vector of the \textit{k}-th synset of candidates
$$
\mean{cand\ syn_k}(\varepsilon)=\frac{1}{s_k(\varepsilon)}\sum_{u\in {cand\ syn_k(\varepsilon)}}u.
$$
If $s_k(\varepsilon)=0$, then $\mean{cand\ syn_k}(\varepsilon)$ equals to the zero vector.

Line \ref{alg2line:similarity}. Calculate the similarity of the average vectors of the filtered sentence and the \textit{k}-th filtered synset
$$
sim_k(\varepsilon)=sim(\mean{{cand}_kS}(\varepsilon), \mean{cand\ syn_k}(\varepsilon)).
$$

Lines \ref{alg2line:sim_max}--\ref{alg2line:unique}. Suppose $max_{k=1, \dots ,l}\{sim_k(\varepsilon)\}=sim_{\scalebox{0.9}{$k^*$}}(\varepsilon)$, i.e. $k^*\in \{1,...,l\}$ is the number of the
largest  $sim_k(\varepsilon)$. 
If $k^*$ is not unique then take another $\varepsilon >0$ 
and repeat the procedure from line \ref{alg2line:epsilon}.

Result: the target word $w^*$ in the sentence $S$ has the sense corresponding to the $k^*$-th synset $syn_{k^*}^w$.

This algorithm was implemented in Python.\footnote{
See the function \textit{selectSynsetForSentenceByAverageSimilarityModified} in the file \url{https://github.com/componavt/wcorpus.py/blob/master/src/test\_synset\_for\_sentence/lib\_sfors/synset\_selector.py}}

\bfullwidth
\begin{algorithm*}[H]
    \caption{Average algorithm with sentence and synonyms $\varepsilon$-filtration ($\mean{A}_\varepsilon$)}
    \label{alg:aver_eps}
\DontPrintSemicolon
\SetAlgoLined
    \KwData{$v^*$ -- vector of the target word $w^*$ with $l$ senses (synsets), \hfill \break
        $v_i \in S$, $S$ -- sentence with the target word $w^*$, $v^* \in S$, \hfill \break
        $\{syn_k\}$ -- synsets of the target word, that is $syn_k \ni v^*$, $k = \overline{1,l}$.
       }
\KwResult{$k^*\in \{1,\dots,l\}$ is the number of the sense of the word $w^*$ in the sentence $S$.}
    \Do{$k^*$ is not unique}{
        take $\varepsilon>0$ \label{alg2line:epsilon}\;
        \BlankLine
        \Comment{foreach synset of the target word}
        \ForEach{$syn_k \ni v^*$}{%
            \Comment{construct the set of words of the sentence $S$ filtered by synonyms of the \textit{k}-th synset $syn_k$:}
            $cand_kS(\varepsilon) = \{v \in S: \exists u\in syn_k, sim(v, u)>\varepsilon, v \neq v^*, u \neq v^*\}$ \;
            $S_k(\varepsilon) = |cand_kS(\varepsilon)|$, \label{alg2line:sentence_size} \Comment{number of candidates of the sentence;}
            \Comment{the average vector of sentence candidates:}
            $\mean{{cand}_kS}(\varepsilon)= 
                \begin{cases}
                    \frac{1}{S_k(\varepsilon)}\sum\limits_{v\in cand_kS(\varepsilon)}v,& \text{\small if $S_k(\varepsilon)>0$} \\
                    \overrightarrow{0},& \text{\small if $S_k(\varepsilon)=0$}
                \end{cases}
                $ \label{alg2line:average_sentence} \;
            \BlankLine
            \Comment{$\varepsilon$-filtration of the synset $syn_k$ by the sentence $S$:}
            $cand\ syn_k(\varepsilon)=\{u\in syn_k: \exists v\in S, sim(u, v)>\varepsilon, u \neq v^*, v \neq v^* \}$ \label{alg2line:synset_filtered} \;
            $s_k(\varepsilon) = |cand\ syn_k(\varepsilon)|$, \label{alg2line:set_size} \Comment{number of candidates of synonyms}
            \Comment{the average vector of synset candidates:}
            $\mean{cand\ syn_k}(\varepsilon)= 
                \begin{cases}
                    \frac{1}{s_k(\varepsilon)}\sum\limits_{u\in cand\ syn_k(\varepsilon)}u,& \text{\small if $s_k(\varepsilon)>0$} \\
                    \overrightarrow{0},& \text{\small if $s_k(\varepsilon)=0$}
                \end{cases}
                $ \label{alg2line:average_synset} \;
            \Comment{the similarity of the average vectors of the sentence and the \textit{k}-th filtered synset:}
            $sim_k(\varepsilon)=sim(\mean{{cand}_kS}(\varepsilon), \mean{cand\ syn_k}(\varepsilon))$ \label{alg2line:similarity} \;
        }
        $sim_{\scalebox{0.9}{$k^*$}}(\varepsilon) = \max_{k=1,\dots,l} \{sim_k(\varepsilon)\} \Rightarrow k^*\in \{1,\dots,l\}$ \label{alg2line:sim_max}, \Comment{$k^*$ is the number of the largest $sim_k(\varepsilon)$}
    } \label{alg2line:unique}
\end{algorithm*}
\efullwidth


\section{$K$-algorithm based on $\varepsilon$-dilatation}

The algorithm~\ref{alg:dilatation} ($K$-algorithm) is based on the function $\tilde{K}(A, B, \varepsilon)$ 
(see previous section ``New $\varepsilon$-proximity between finite sets'' on page~\pageref{Section:NewEpsProximity}), 
where  $A=syn_k$, that is \textit{k}-th synset, and $B=S$, where $S$ is a sentence. 
The algorithm includes the following steps.

Lines \ref{alg3line:epsilon}--\ref{alg3line:near_set}. 
Given $\varepsilon >0$, let us construct 
the $C_k(\varepsilon)$ set of ``near'' words of the \textit{k}-th synset and the sentence~$S$.

Line \ref{alg3line:distant_set}. 
Denote by $D_k(\varepsilon)$ the set of ``distant'' words 
$$ D_k(\varepsilon)=(S\cup syn_k)\setminus C_k(\varepsilon).$$

Line \ref{alg3line:ratio}. 
Calculate $\tilde{K}_k(\varepsilon)$ as the ratio of ``near'' and ``distant'' elements of the sets 
$$
\tilde{K}_k(\varepsilon)=\frac{|C_k(\varepsilon)|}{1+|D_k(\varepsilon)|}.
$$

Lines \ref{alg3line:max}--\ref{alg3line:unique}. 
Suppose ${max_{k=1,...,l}\tilde{K}_k(\varepsilon)=\tilde{K}_{k^*}(\varepsilon)}$. If $k^*$ is not unique, 
then take another $\varepsilon>0$ 
and repeat the procedure from line~\ref{alg3line:epsilon}.


{\centering
\begin{minipage}{1\linewidth}
  \begin{algorithm}[H]

    \caption{$K$-algorithm based on $\varepsilon$\=/dilatation}
    \label{alg:dilatation}
\DontPrintSemicolon
\SetAlgoLined
    \KwData{$v^*$ -- vector of target word $w^*$ with $l$~senses (synsets),
        $v_i \in S$, $v^* \in S$, \hfill \break
        $\{syn_k\}$ -- synsets of $v^*$, $k = \overline{1,l}$.
       }
\KwResult{$k^*\in \{1,\dots,l\}$ is the number of the sense of the word $w^*$ in the sentence $S$.}
\BlankLine
    \Do{$k^*$ is not unique}{
        take $\varepsilon>0$ \label{alg3line:epsilon}\;
        \BlankLine
        \Comment{foreach synset of the target word}
        \ForEach{$syn_k \ni v^*$}{%
            \Comment{set of near words:}
            $C_k(\varepsilon)=\{u, v:  {u\in syn_k}, {v\in S}, sim(u, v)>\varepsilon\}$ \label{alg3line:near_set} \;
            \Comment{set of distant words:}
            $D_k(\varepsilon)=(S\cup syn_k)\setminus C_k(\varepsilon)$ \label{alg3line:distant_set} \;
            \Comment{ratio of ``near'' and ``distant'' elements of the sets:}
            $\tilde{K}_k(\varepsilon)=\frac{|C_k(\varepsilon)|}{1+|D_k(\varepsilon)|}$ \label{alg3line:ratio} \;
        }
        \BlankLine
        \Comment{get the number of the largest ratio $k^*$}
        $\tilde{K}_{\scalebox{0.9}{$k^*$}}(\varepsilon) = \max_{k=1,\dots,l} \tilde{K}_k(\varepsilon)$ \label{alg3line:max}
    } \label{alg3line:unique}

  \end{algorithm}
\end{minipage}
\par
}


Result: the target word $w^*$ has the sense corresponding to the $k^*$-th synset $syn_{k^*}^w$.

An example of constructing C and D sets is presented in Fig.~\ref{fig:Isoline} 
and Table. 
It uses the same source data as for the $\mean{A}_0$-algorithm, see Fig.~\ref{fig:VerticesSource}.

Remark. This algorithm is applicable to the $K$-function described in the previous section\footnotemark[\value{footnote}] as well. 
This algorithm was implemented in Python.\footnote{
See the function \textit{selectSynsetForSentenceByAlienDegree} in the file \url{https://github.com/componavt/wcorpus.py/blob/master/src/test\_synset\_for\_sentence/lib\_sfors/synset\_selector.py}}

More details for this example (Fig.~\ref{fig:Isoline}) are presented 
in Table, 
which shows $C$ and $D$ sets with different $\varepsilon$ and values of the $\tilde{K}$-function. 

%
Bold type of word-vertices in Table 
indicates new vertices. 
These new vertices are captured by a set of ``near'' vertices $C$ and these vertices are excluded from the set of ``distant'' vertices $D$ with each subsequent dilatation extension 
with each subsequent $\varepsilon$.
%
%
For example, in the transition from $\varepsilon_1$ to $\varepsilon_2$ 
the set $D_2 (\varepsilon_1)$ loses the vertex $v_3$. 
During this transition ${\varepsilon_1 \rightarrow \varepsilon_2}$ 
the set $C_2 (\varepsilon_2)$ gets the same vertex $v_3$ 
in comparison with the set $C_2 (\varepsilon_1)$.

%
In Fig.~\ref{fig:StepFunctions}, the function $\tilde{K}_1(\varepsilon)$ 
shows the proximity of the sentence $S$ and the synset $syn_1$, 
the function $\tilde{K}_2(\varepsilon)$ -- the proximity of $S$ and the synset $syn_2$.
%
%
It can be seen in Figure~\ref{fig:StepFunctions} that with decreasing $\varepsilon$, the value of $\tilde{K}_2(\varepsilon)$ grows faster than $\tilde{K}_1(\varepsilon)$.

%
Therefore, the sentence $S$ is closer to the second synset $syn_2$. 
The same result can be seen in the previous Fig.~\ref{fig:Isoline}.

\bfullwidth
\efullwidth








\begin{figure}[H]
   \centering

\ifmonochrome
\else
    \includegraphics[keepaspectratio=true,width=0.99\columnwidth]{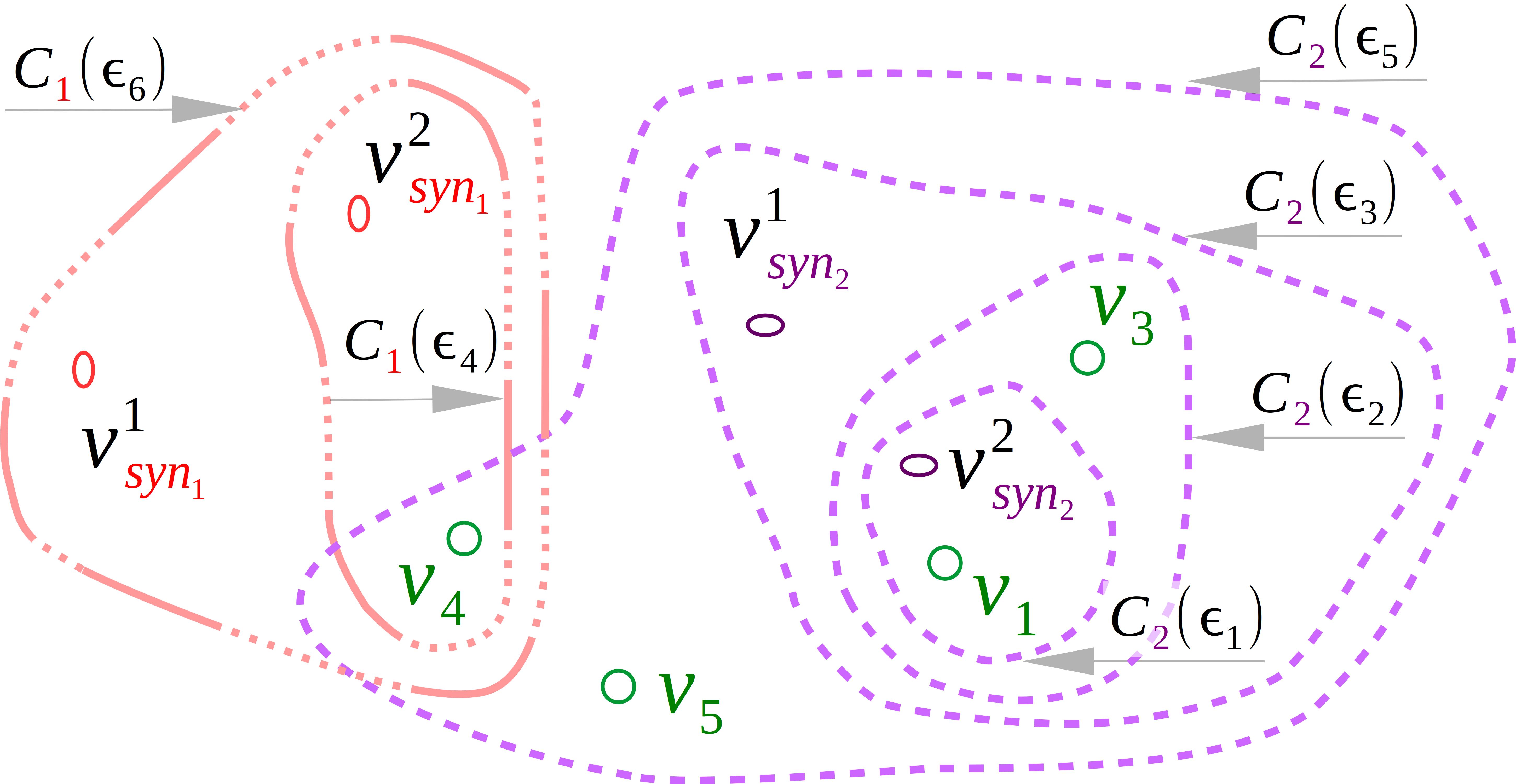}
\fi
   

    \caption{An example of series of $C_k(\varepsilon)$ (sets of words of \textit{k}-th synset 
    which are close and near to the sentence $S$) in the $K$-algorithm based on $\varepsilon$-dilatation. 
    The growth of the dilation of the vertices of the second synset
$\{v^1_{syn_2}, v^2_{syn_2}\}$ 
captures the vertices of the sentence $S = \{v_1, v_3, v_4, v_5\}$ faster than the dilation of the vertices of the first synset. In other symbols: 
$(syn_2 + \varepsilon) \cap S \, \supset \, (syn_1 + \varepsilon) \cap S$. 
That is, according to the $K$-algorithm, the second value of the word-vector $v_2$, represented by the synset $syn_2$, will be selected for the sentence $S$}
    \label{fig:Isoline}
\end{figure}

\begin{figure}[H]
   \centering
\ifmonochrome
\else
    \includegraphics[keepaspectratio=true,width=0.99\columnwidth]{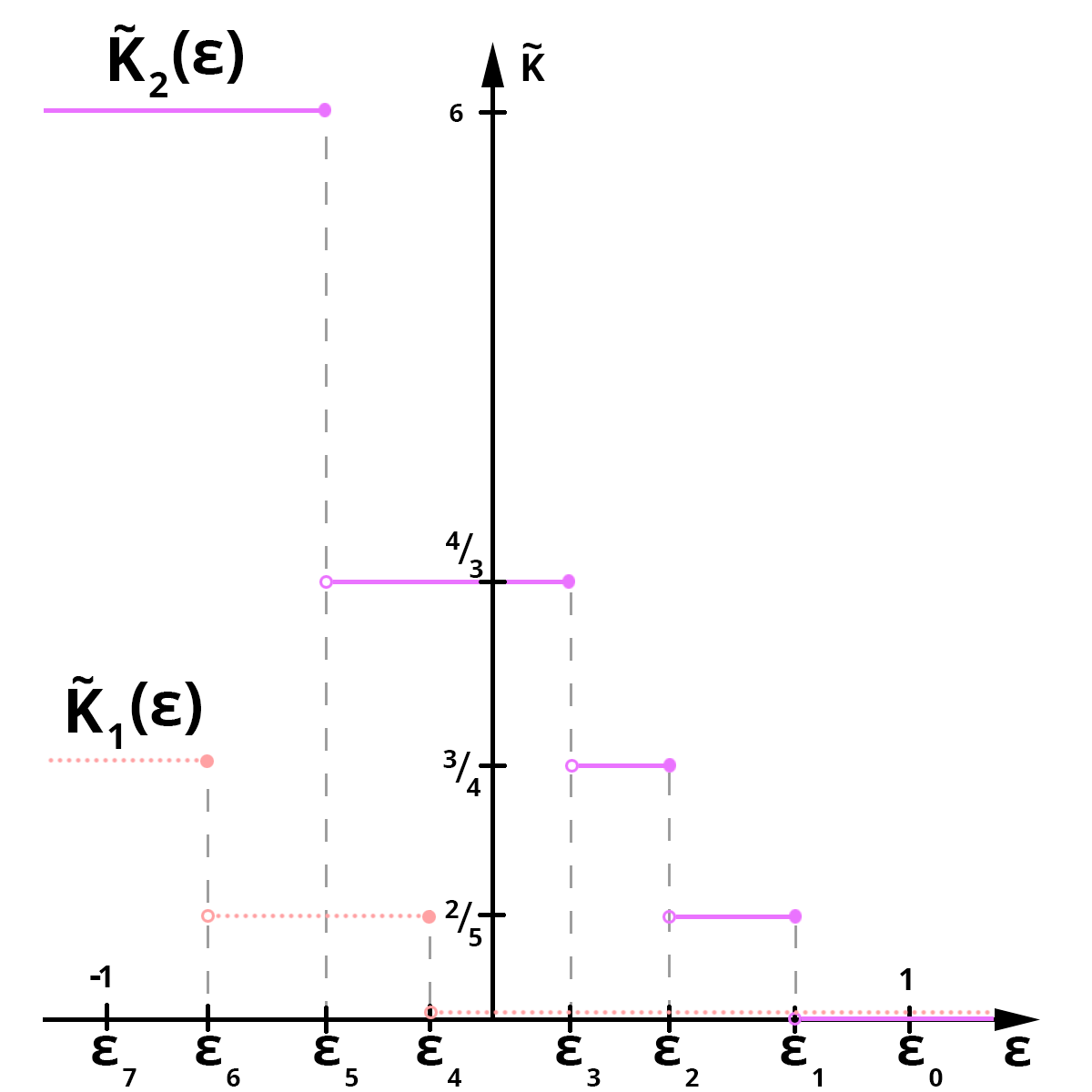}
\fi

    \caption{Left-continuous step functions $\tilde{K}_1(\varepsilon)$, $\tilde{K}_2(\varepsilon)$ show that the sentence $S$ is closer to the second synset $syn_2$}
    \label{fig:StepFunctions}
\end{figure}


\bfullwidth
\begin{table}[H]
\centering 
    \caption*{
    An example of the $K$-algorithm treating the word $w_2$, which has two synsets    
    $syn_1$, $syn_2$ and the sentence $S$, where $w_2 \in S$, see Fig.~\ref{fig:WiktSlushit}. 
    The number of the algorithm iteration corresponds to the index of $\varepsilon$. 
    Let the series of $\varepsilon$ be ordered so that 
    $1=\varepsilon_0 > \varepsilon_1 > \varepsilon_2 > ... > \varepsilon_7=-1$.
    It is known that $|C_1 \cup D_1 \setminus v_2| = |S \setminus v_2| = 6$, that is the total number of words in the synsets and in the sentence are constants.
    }
    \label{tab:NativeAndAlienSets}
    \begin{tabular}{ c l c c c c }
\hline
$\epsilon$ & \CIIepsilon & $D_2(\epsilon)$ & $|C_2|$ & $|D_2|$ & $\tilde{K}_2(\epsilon)$ \\
\hline
    \multicolumn{6}{r}{$\tilde{K}_k(\epsilon)=\frac{|C_k(\epsilon)|}{1+|D_k(\epsilon)|}$} \\ \hline

        $\epsilon_0$ & $\varnothing$ & \vbI, $v_3$, $v_4$, $v_5$, $v^1_{syn_2}$, \vbIIsynII & 0 & 6 & 0.0 \\ \hline 
    
    $\epsilon_1$ & \vI, \vIIsynII & \vbIII, $v_4$, $v_5$, \vIsynII & 2 & 4 & $\frac{2}{5}$ \\ \hline
    
    $\epsilon_2$ & \vI, \vIIsynII, \vbIII & $v_4$, $v_5$, \vbIsynII & 3 & 3 & $\frac{3}{4}$ \\ \hline
    
    $\epsilon_3$ & \vI, \vIIsynII, \vIII, \vbIsynII & \vbIV, \vbV & 4 & 2 & $\frac{4}{3}$ \\ \hline

    
\hline
     & \CIepsilon & $D_1(\epsilon)$ & $|C_1|$ & $|D_1|$ & $\tilde{K}_1(\epsilon)$ \\
\hline
    
    $\epsilon_4$ & \vIIsynI, \vIV & \vbIsynI, $v_1$, $v_3$, $v_5$ & 2 & 4 & $\frac{2}{5}$ \\ \hline
    
\hline
     & \CIIepsilon & $D_2(\epsilon)$ & $|C_2|$ & $|D_2|$ & $\tilde{K}_2(\epsilon)$ \\
\hline
    
    $\epsilon_5$ & \vI, \vIIsynII, \vIII, \vIsynII, \vbIV, \vbV, & $\varnothing$ & 6 & 0 & 6 \\ \hline

\hline
     & \CIepsilon & $D_1(\epsilon)$ & $|C_1|$ & $|D_1|$ & $\tilde{K}_1(\epsilon)$ \\
\hline

    $\epsilon_6$ & \vIIsynI, \vIV, \vbIsynI & $v_1$, $v_3$, $v_5$ & 3 & 3 & $\frac{3}{4}$ \\ \hline
    
    \end{tabular}
\end{table}
\efullwidth

\section{Experiments}

\subsection{Web of tools and resources} \label{Section:WebOfTools}

This section describes the resources used in our research, namely: Wikisource, Wiktionary, WCorpus and RusVectores.

The developed WCorpus\footnote{\url{https://github.com/componavt/wcorpus}} system includes texts extracted from Wikisource and provides 
the user with a text corpus analysis tool. 
This system is based on the Laravel framework (PHP programming language). MySQL database is used.\footnote{See WCorpus database scheme: \url{https://github.com/componavt/wcorpus/blob/master/doc/workbench/db\_scheme.png}}

\textit{Wikisource}. The texts of Wikipedia have been used as a basis for several contemporary corpora~\cite{Jurczyk2018}. 
But there is no mention of using texts from Wikisource in text processing. Wikisource is an open online digital library with texts in many languages. Wikisource sites contains 10 millions of texts\footnote{\url{https://stats.wikimedia.org/wikisource/EN/TablesWikipediaZZ.htm}} in more than 38 languages.\footnote{\url{https://stats.wikimedia.org/wikisource/EN/Sitemap.htm}} Russian Wikisource 
(the database dump as of February 2017) was used in our research.

\textit{Texts parsing}. The texts of Wikisource were parsed, analysed and stored to the WCorpus database. Let us describe this process in detail. The database dump containing all texts of Russian Wikisource was taken from ``Wikimedia Downloads'' site.\footnote{\url{https://dumps.wikimedia.org/backup-index.html}} These Wikisource database files were imported into the local MySQL database titled ``Wikisource Database'' 
in Fig.~\ref{fig:WebOfToolsArch}, where ``WCorpus Parser'' is the set of WCorpus PHP-scripts which analyse and parse the texts in the following three steps.

\begin{enumerate}
\item First, the title and the text of an article from the Wikisource database are extracted (560 thousands of texts). 
    One text corresponds to one page on Wikisource site. 
    It may be small (for example, several lines of a poem), medium (chapter or short story), 
        or huge size (e.g. the size of the page with the novella ``The Eternal Husband'' written by Fyodor Dostoyevsky is 500 KB). 
Text preprocessing includes the following steps:
\begin{itemize}
\item     Texts written in English and texts in Russian orthography before 1918 were excluded; about 12 thousands texts were excluded. 
\item     Service information (wiki markup, references, categories and so on) was removed from the text.
\item     Very short texts were excluded. As a result, 377 thousand texts were extracted.
\item     Texts splitting into sentences produced 6 millions of sentences.
\item     Sentences were split into words (1.5 millions of unique words).
\end{itemize}
\end{enumerate}

\bfullwidth
\begin{figure}[h!tb]
   \centering
\ifmonochrome
\else
    \includegraphics[keepaspectratio=true,width=0.65\columnwidth]{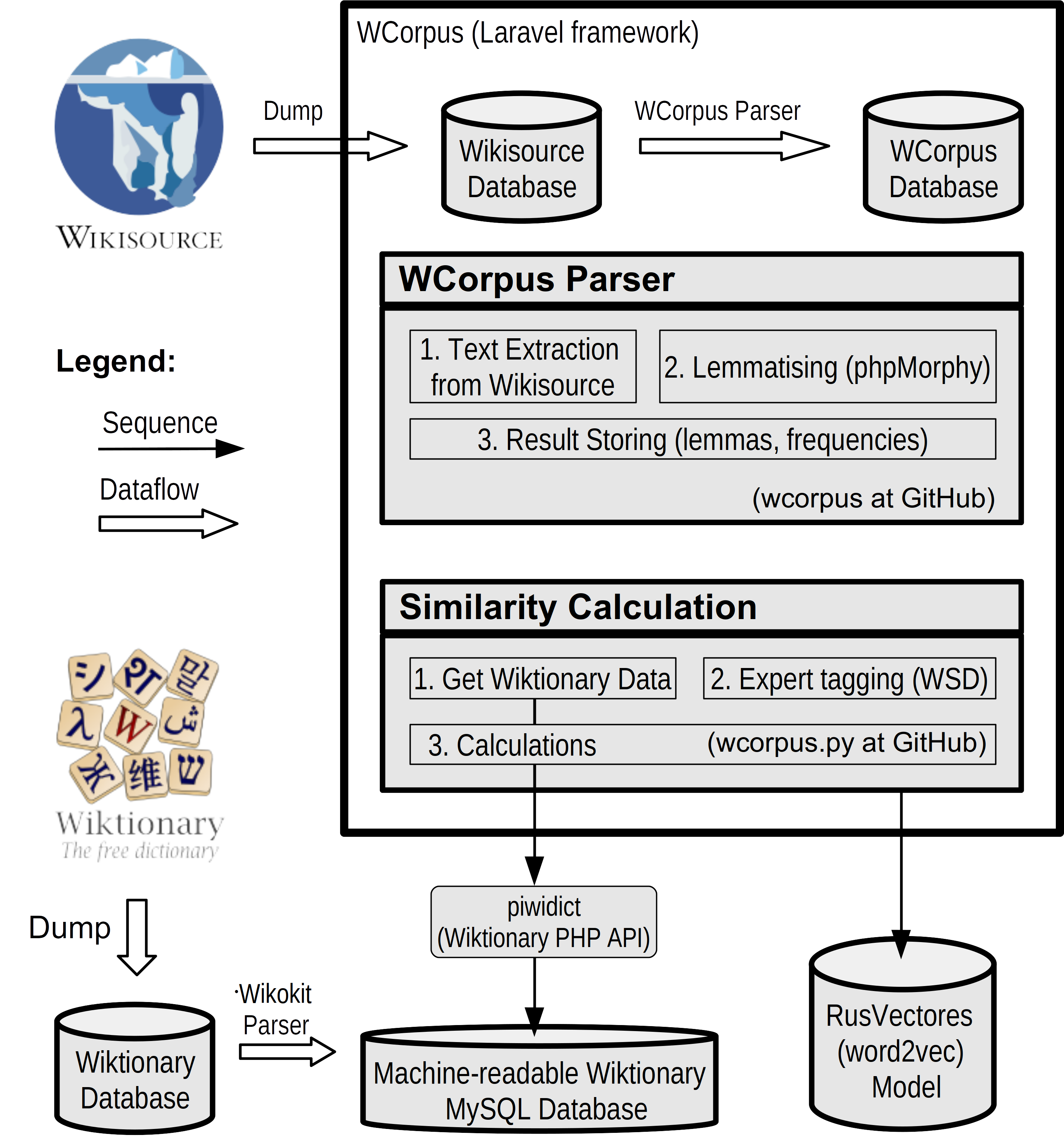}
\fi
    \caption{The architecture of WCorpus system and the use of other resources}
    \label{fig:WebOfToolsArch}
\end{figure}
\efullwidth

\begin{enumerate}
  \setcounter{enumi}{2}
\item Secondly, word forms were lemmatized using phpMorphy\footnote{\url{https://packagist.org/packages/componavt/phpmorphy}} program (0.9 million lemmas).
\item Lastly, lemmas, wordforms, sentences and relations between words and sentences were stored to WCorpus database (Fig.~\ref{fig:WebOfToolsArch}).
\end{enumerate}

In our previous work ``Calculated attributes of synonym sets''~\cite{krizhanovsky2018calculated} we also used neural network models of the great project RusVectores\footnote{\url{http://rusvectores.org/en/}}, which is a kind of a word2vec tool based on Russian texts~\cite{kutuzov2015texts}.

\subsection{Context similarity algorithms evaluation}

In order to evaluate the proposed WSD algorithms, 
several words were selected from a dictionary, 
then sentences with these words were extracted from the corpus and tagged by experts.

\vfill\null
\columnbreak

\subsubsection{Nine words}

Only polysemous words which have at least two meanings with different sets of synonyms are suitable for our evaluation of WSD algorithms.

The following criteria for the selection of synonyms and sets of synonyms from Russian Wiktionary were used:
\begin{enumerate}
\item Only single-word synonyms are extracted from Wiktionary. This is due to the fact that the RusVectores neural network model ``ruscorpora\_2017\_1\_600\_2'' used in our research does not support multiword expressions.

\item If a word has meanings with equal sets of synonyms, then these sets were skipped 
    because it is not possible to discern different meanings of the word using only these synonyms without additional information.
\end{enumerate}

A list of polysemous words was extracted from the parsed Russian Wiktionary\footnote{\url{http://whinger.krc.karelia.ru/soft/wikokit/index.html}} using PHP API piwidict\footnote{\url{https://github.com/componavt/piwidict}} (Fig.~\ref{fig:WebOfToolsArch}).

Thus, 9 polysemous Russian words (presented in the subcorpus\footnote{See information about the subcorpus in the section ``Sentences of three Russian writers'' on page~\pageref{Section:3writers}.}) were selected by experts from this Wiktionary list, namely: ``бездна'' (bezdna), ``бросать'' (brosat'), ``видный'' (vidnyy), ``донести'' (donesti), ``доносить'' (donosit'), ``занятие'' (zanyatiye), ``лихой'' (likhoy), ``отсюда'' (otsyuda), ``удачно'' (udachno). The tenth word ``служить'' (sluzhit') was left out of consideration, because there are 1259 of 1308 sentences with this frequent word to be tagged by experts in the future (Fig.~\ref{fig:Slushit}).

\bfullwidth
\begin{figure}[H]
   \centering
\ifmonochrome
\else
    \includegraphics[keepaspectratio=true,width=0.82\columnwidth]{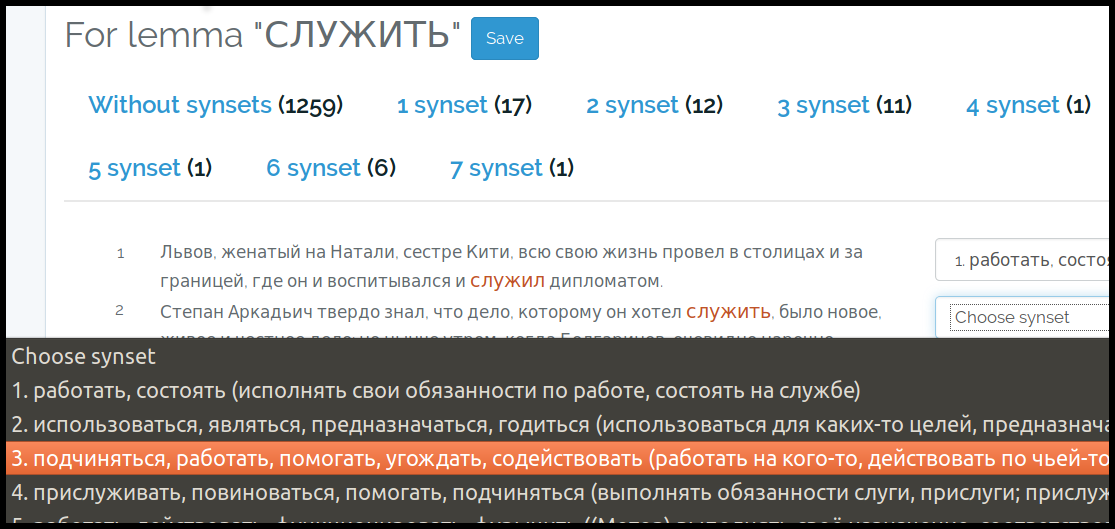}
\fi
    \caption{Russian verb ``служить'' (sluzhit') has seven meanings and seven synsets 
    in the developed system WCorpus. 
    49 sentences are already linked to relevant senses of this verb. 
    1259 sentences remain to be tagged by experts}
    \label{fig:Slushit}
\end{figure}
\efullwidth

\subsubsection{Sentences of three Russian writers} \label{Section:3writers}

The sentences which contain previously defined 9 words were to be selected from the corpus and tagged by experts. 
But the Wikisource corpus was too huge for this purpose.  
So, in our research a subcorpus of Wikisource texts was used. These are the texts written by Fyodor Dostoevsky, Leo Tolstoy and Anton Chekhov.

Analysis of the created WCorpus database with texts of three writers shows that the subcorpus contains:\footnote{See SQL-queries applied to the subcorpus \url{https://github.com/componavt/wcorpus/wiki/SQL}}
\begin{itemize}
\item 2635 texts;
\item 333 thousand sentences;
\item 215 thousand wordforms;
\item 76 thousand lemmas;
\item 4.3 million wordform-sentence links;
\end{itemize}

Texts of this subcorpus contain 1285 sentences with these 9 words, 
wherein 9 words have in total 42 synsets (senses). 
It was developed 
A graphical user interface (webform) of the WCorpus system (Fig.~\ref{fig:Slushit}) was developed, 
where experts selected one of the senses of the target word for each of the 1285 sentences. 

This subcorpus database with tagged sentences and linked synsets is available online~\cite{writers}.

\subsubsection{Text processing and calculations}

These 1285 sentences were extracted from the corpus. Sentences were split into tokens. Then wordforms were extracted. 
All the wordforms were lowercase and lemmatized. 
Therefore, a~sentence is a bag of words. Sentences with only one word were skipped.

The phpMorpy lemmatizer takes a wordform and yields possible lemmas with the corresponding part of speech (POS). 
Information on POS of a word is needed to work with the RusVectores' 
prediction neural network model ``ruscorpora\_2017\_1\_600\_2'', 
because to get a vector it is necessary to ask for a word and POS, for example ``serve\_VERB''. Only nouns, verbs, adjectives and adverbs remain in a sentence bag of words, other words were skipped.

The computer program (Python scripts) which works with the WCorpus database and RusVectores was written and presented in the form of the project \textit{wcorpus.py} at GitHub.\footnote{\url{https://github.com/componavt/wcorpus.py}} The source code in the file \textit{synset\_selector.py}\footnote{\url{https://github.com/componavt/wcorpus.py/blob/master/src/test\_synset\_for\_sentence/lib\_sfors/synset\_selector.py}} implements three algorithms described in the article, namely:
\begin{itemize}
\item $\mean{A}_0$-algorithm implemented in the function \textit{selectSynsetForSentenceByAverage\-Si\-mi\-la\-ri\-ty()};
\item $K$-algorithm -- function \textit{selectSynsetForSen\-ten\-ce\-By\-AlienDegree()};
\item $\mean{A}_\varepsilon$-algorithm -- function \textit{selectSynsetForSen\-tence\-By\-Ave\-rageSimilarityModified()}.
\end{itemize}

These three algorithms calculated and selected one of the possible synsets for each of 1285 sentences.

Two algorithms ($K$ and $\mean{A}_\varepsilon$) have an input parameter of $\varepsilon$, therefore, 
a cycle with a step of 0.01 from 0 to 1 was added, which resulted in 100 iterations for each sentence.

Then, answers generated by the algorithms were compared with the synsets selected by experts.

The number of sentences with the sense correctly tagged by the $K$-algorithm for nine Russian words presented in Fig.~\ref{fig:NotKandinsky}.

The legend of this figure lists target words with numbers in brackets $(X, Y)$, 
where $X$ is the number of sentences with these words, $Y$ is the number of senses.


\ifmonochrome
\else 
The curves for the words ``ЗАНЯТИЕ'' (``ZANYATIYE'', cyan solid line with star points) and ``ОТСЮДА'' (``OTSYUDA'', green solid line with triangle points) are quite high for some $\varepsilon$, because (1) there are many sentences with these words (352 and 308) in our subcorpus, (2) these words have few meanings (3 and 2).
\fi 

\bfullwidth
\begin{figure}[H]
   \centering

\ifmonochrome
\else
    \includegraphics[keepaspectratio=true,width=0.97\columnwidth]{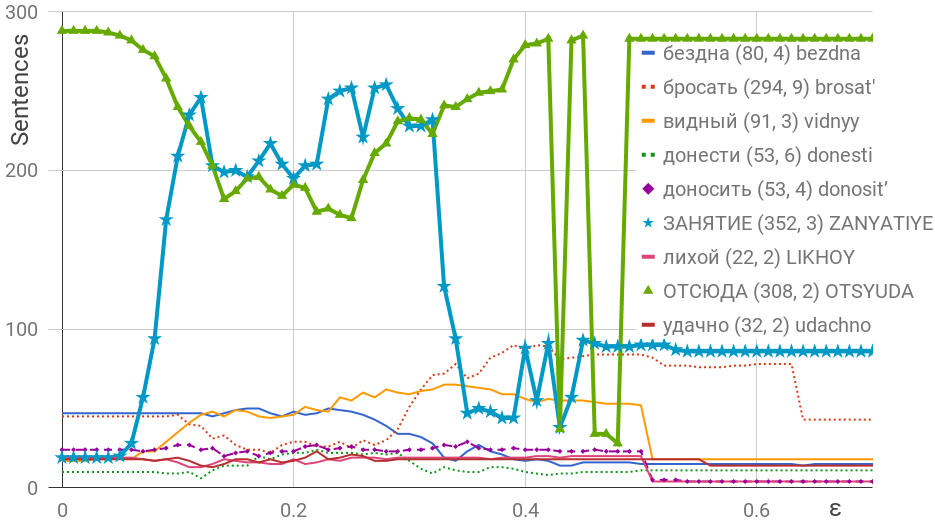}
\fi

    \caption{Number of sentences with the correct tagged sense for nine Russian words by the \textit{K}-algorithm}
    \label{fig:NotKandinsky}
\end{figure}

\begin{figure}[H]
   \centering

\ifmonochrome
\else
    \includegraphics[keepaspectratio=true,width=0.97\columnwidth]{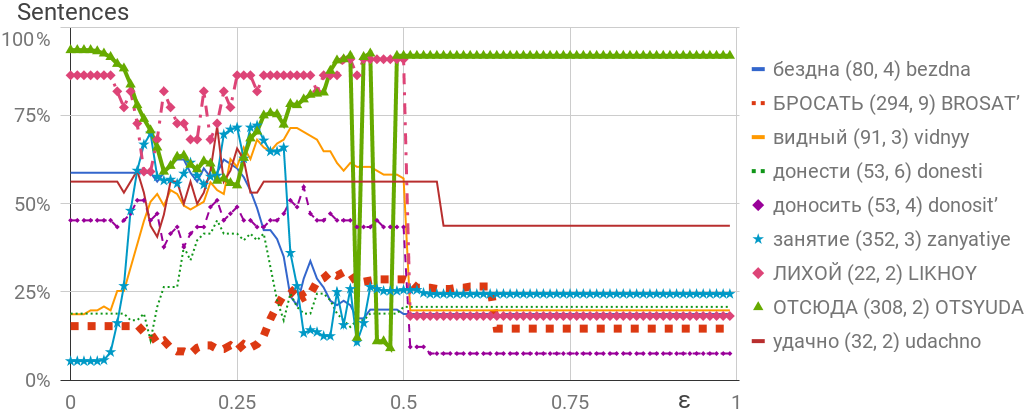}
\fi

    \caption{Normalised data with the fraction of sentences with correctly tagged sense for nine Russian words}
    \label{fig:NotKandinskyNormalised}
\end{figure}
\efullwidth

\textit{More meanings, poorer results.}

\ifmonochrome
\else 
If a word has more meanings, then the algorithm yields even poorer results. It is visible in the normalised data (Fig.~\ref{fig:NotKandinskyNormalised}), where examples with good results are ``ОТСЮДА'' (OTSYUDA) and ``ЛИХОЙ'' (LIKHOY, pink dash dot line with diamond points) with 2 meanings; the example ``БРОСАТЬ'' (BROSAT', red bold dotted line) with 9 meanings has the worst result (the lowest dotted curve).
\fi

\subsection{Comparison of three algorithms}

Let us compare three algorithms by summing the results for all nine words. 
Fig.~\ref{fig:3alg} contains the following curves:
\ifmonochrome
\else 
    \begin{itemize}
    \item $\mean{A}_0$-algorithm -- long dash blue line;
    \item $K$-algorithm -- solid red line;
    \item $\mean{A}_\varepsilon$-algorithm -- dash yellow line.
    \end{itemize}
\fi

The $\mean{A}_0$-algorithm does not depend on $\varepsilon$. It showed mediocre results.

The $K$\=/algorithm yields better results than $\mean{A}_\epsilon$\=/algorithm when $\varepsilon > 0.15$.

The $K$\=/algorithm showed the best results on the interval [0.15; 0.35]. Namely, more than 700 sentences (out of 1285 human-tagged sentences) were properly tagged with the $K$\=/algorithm on this interval (Fig.~\ref{fig:3alg}).

\bfullwidth
\begin{figure}[H]
   \centering

\ifmonochrome
\else
    \includegraphics[keepaspectratio=true,width=0.67\columnwidth]{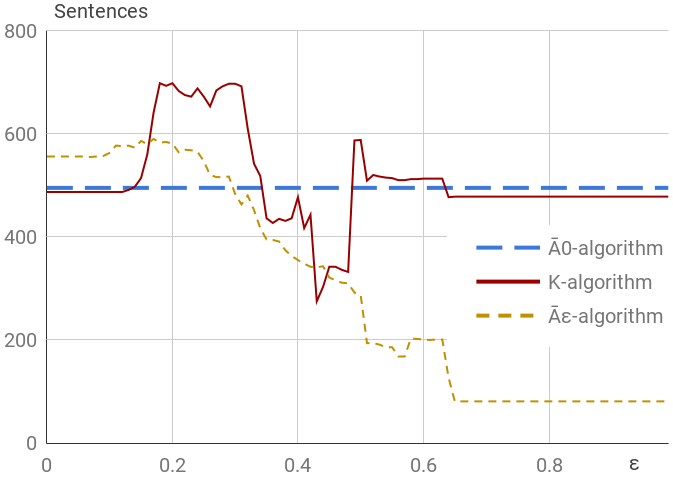}
\fi

    \caption{Comparison of $\mean{A}_0$-algorithm, $K$\=/algorithm, $\mean{A}_\varepsilon$-algorithm}
    \label{fig:3alg}
\end{figure}
\efullwidth

\subsection{Comparison of four algorithms as applied to nine words}

Let us compare the results of running four algorithms for each word separately (Fig.~\ref{fig:4alg}):
\ifmonochrome
\else
    \begin{itemize}

    \item $\mean{A}_0$-algorithm -- long dash blue line with triangle points;

    \item $K$-algorithm -- solid red line with square points;

    \item $\mean{A}_\varepsilon$-algorithm -- dash yellow line with circle points;

    \item ``Most frequent meaning'' -- green dashed line with X marks.
    \end{itemize}
\fi

The simple ``most frequent meaning'' algorithm was added to compare the results. 
This algorithm does not depend on the variable~$\varepsilon$, 
it selects the meaning (synset) that is the most frequent in our corpus of texts. 
In Fig.~\ref{fig:4alg} 
\ifmonochrome
\else
this algorithm corresponds to a green dashed line with X marks.
\fi

The results of the ``most frequent meaning'' algorithm and $\mean{A}_0$-algorithm are similar (Fig.~\ref{fig:4alg}). 

The $K$-algorithm is the absolute champion in this competition, 
that is for each word there exists an $\varepsilon$ such that the $K$\=/algorithm outperforms other algorithms (Fig.~\ref{fig:4alg}).

Let us explain the calculation of the curves in Fig.~\ref{fig:4alg}.

For the $\mean{A}_0$-algorithm and the ``most frequent meaning'' algorithm, 
the meaning (synset) is calculated for each of the nine words on the set of 1285 sentences. 
Thus, $1285 \cdot 2$ calculations were performed.

And again, the $\mean{A}_\varepsilon$-algorithm and the $K$\=/algorithm depend on the variable $\varepsilon$. 
But how can the results be shown without the $\varepsilon$ axis? 
If at least one value of $\varepsilon$ gives a positive result, 
then we suppose that the WSD problem was correctly solved for this sentence by the algorithm.

%
The value on the Y axis for the selected word (for $\mean{A}_\varepsilon$-algorithm and $K$\=/algorithm) is equal to the sum of such correctly determined sentences (with different values of $\varepsilon$) in Fig.~\ref{fig:4alg}.

\bfullwidth
\begin{figure}[H]
   \centering
\ifmonochrome
\else
    \includegraphics[keepaspectratio=true,width=0.82\columnwidth]{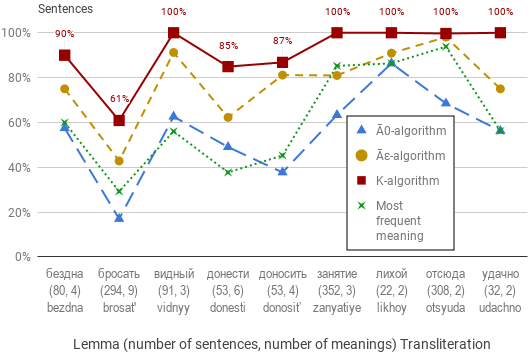}
\fi
    \caption{Comparison of $\mean{A}_0$-algorithm, $K$\=/algorithm, $\mean{A}_\varepsilon$-algorithm and the most frequent meaning}
    \label{fig:4alg}
\end{figure}
\efullwidth

%
Perhaps it would be more correct 
to fix $\varepsilon$ corresponding to the maximum number of correctly determined sentences. 
Then, the result will not be so optimistic.

%
To show the complexity of comparing and evaluating $\varepsilon$-algorithms (that is, algorithms that depend on $\varepsilon$), let us try to analyze the results of the $K$\=/algorithm, shown in Fig~\ref{fig:OneDistributed}.

%
The percentage (proportion) of correctly determined 1285 sentences for 9 words by the $K$\=/algorithm, where the $\varepsilon$ variable changes from 0 to 1 in increments of 0.01, is presented in Fig.~\ref{fig:OneDistributed}. Thus, $1285 \cdot 100$ calculations were performed.

%
These proportions are distributed over a set of possible calculated results from 0\% (no sentence is guessed) to 100\% (all sentences are guessed) for each of nine words.

%
This Figure~\ref{fig:OneDistributed} does not show which $\varepsilon$-values 
produce better or poorer results, although it could be seen in Figures \ref{fig:NotKandinsky}--\ref{fig:3alg}. 
But the Figure does show the set and the quality of the results obtained with the help of the $K$\=/algorithm. 
For example, the word ``лихой'' (likhoy) with 22 sentences and 100 different $\varepsilon$ has only 8 different outcomes of the $K$\=/algorithm, seven of which lie in the region above 50\%, that is, more than eleven sentences are guessed at any $\varepsilon$.


For example, the word ``бросать'' (brosat') has the largest number of meanings in our data set, it has 9 synonym sets in our dictionary and 11 meanings in Russian Wiktionary.\footnote{\href{https://ru.wiktionary.org/wiki/\%D0\%B1\%D1\%80\%D0\%BE\%D1\%81\%D0\%B0\%D1\%82\%D1\%8C}{https://ru.wiktionary.org/wiki/бросать}}
Аll possible results of the $K$\=/algorithm for this word are distributed in the range of 10--30\%. 
The maximum share of guessed sentences is 30.61\%.
Note that this value is achieved when $\varepsilon = 0.39$, and this is clearly shown in Figure~\ref{fig:NotKandinskyNormalised}, see the thick dotted line.


All calculations, charts drawn from experimental data and results of the experiments are available online in Google Sheets~\cite{GoogleSheetsExperiments}.

\bfullwidth
\begin{figure}[H]
   \centering
\ifmonochrome
\else
    \includegraphics[keepaspectratio=true,width=0.85\columnwidth]{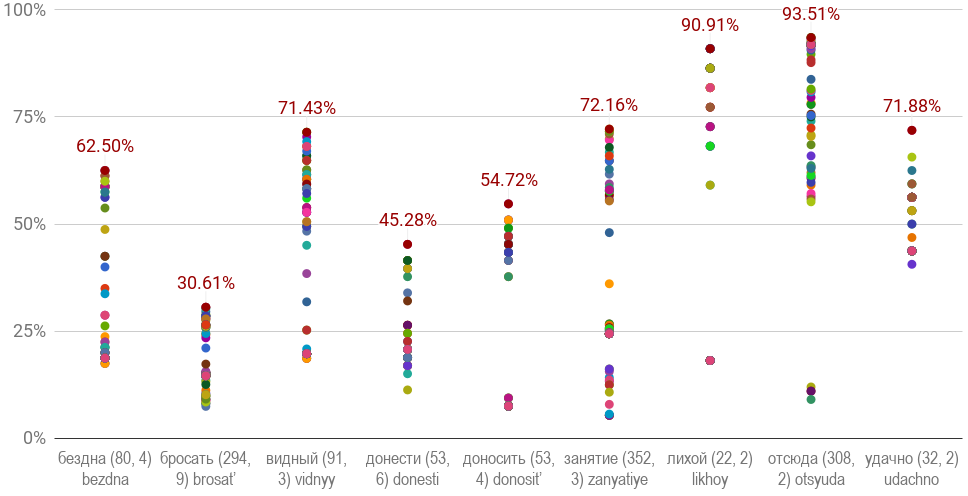}
\fi
    \caption{Proportions of correctly guessed sentences distributed over a set of possible calculated results}
    \label{fig:OneDistributed}
\end{figure}
\efullwidth

\section{Conclusions}

%
The development of the corpus analysis system WCorpus\footnote{\url{https://github.com/componavt/wcorpus}} was started. 
377 thousand texts were extracted from Russian Wikisource, processed and uploaded to this corpus. 

%
Context-predictive models of the RusVectores project are used to calculate the distance between lemmas.
Scripts in Python were developed to process RusVectores data, see the \textit{wcorpus.py} project on the GitHub website.

The WSD algorithm based on a new method of vector-word contexts proximity calculation is proposed and implemented. Experiments have shown that in a number of cases the new algorithm shows better results.  

%
%
The future work is matching Russian lexical resources (Wiktionary, WCorpus) to Wikidata objects~\cite{Nielsen2018}.

\textit{The study was supported by the Russian Foundation for Basic Research, grant \mbox{No.~18-012-00117}}.



\vspace*{0.5cm}
\begin{thebibliographyen}{9}

\bibitem{Arora2017simple}
    \textit{Arora~S., Liang~Y., Ma~T.} A simple but tough-to-beat baseline for sentence embeddings. \textit{In Proceedings of the ICLR}, 2017. P.~1--16. URL: \url{https://pdfs.semanticscholar.org/3fc9/7768dc0b36449ec377d6a4cad8827908d5b4.pdf} (access date: 3.04.2018).

\bibitem{Chen2014unified}
\textit{Chen~X., Liu~Z., Sun~M.}
        A unified model for word sense representation and disambiguation. \textit{In Proceedings of the EMNLP}, 2014. P.~1025--1035. doi: 10.3115/v1/d14-1110. URL: \url{http://www.aclweb.org/anthology/D14-1110} (access date: 3.04.2018). 
        
\bibitem{Choi2010survey}
\textit{Choi~S.~S., Cha~S.~H., Tappert~C.~C.}
        A survey of binary similarity and distance measures. \textit{Journal of Systemics, Cybernetics and Informatics}. 2010. Vol.~8. no.~1. P.~43--48. URL: \url{http://citeseerx.ist.psu.edu/viewdoc/download?doi=10.1.1.352.6123&rep=rep1&type=pdf} (access date: 3.04.2018).

\bibitem{Haussler1999convolution}
\textit{Haussler~D.} 
        Convolution kernels on discrete structures. \textit{Technical report, Department of Computer Science, University of California at Santa Cruz}. 1999. URL: \url{https://www.soe.ucsc.edu/sites/default/files/technical-reports/UCSC-CRL-99-10.pdf} (access date: 3.04.2018).

\bibitem{Jurczyk2018}
\textit{Jurczyk~T., Deshmane~A., Choi~J.} 
Analysis of Wikipedia-based corpora for question answering. 
        \textit{arXiv preprint arXiv:1801.02073}. 2018. URL: \url{http://arxiv.org/abs/1801.02073} (access date: 3.04.2018).

\bibitem{krizhanovsky2018calculated}
\textit{Krizhanovsky~A., Kirillov~A.} 
Calculated attributes of synonym sets. 
        \textit{arXiv preprint arXiv:1803.01580}. 2018. URL: \url{http://arxiv.org/abs/1803.01580} (access date: 3.04.2018).

\bibitem{writers}
\textit{Krizhanovsky~A., Kirillov~A., Krizhanovskaya~N.} 
        WCorpus mysql database with texts of 3 writers. \textit{figshare}. 2018. URL: \url{https://doi.org/10.6084/m9.figshare.5938150.v1} (access date: 3.04.2018).

\bibitem{GoogleSheetsExperiments}
\textit{Krizhanovsky~A., Kirillov~A., Krizhanovskaya~N.} 
        Assign senses to sentences of 3 writers. \textit{Google Sheets}. 2018. URL: \url{http://bit.ly/2I14QIT} (access date: 27.04.2018).

\bibitem{kutuzov2015texts}
\textit{Kutuzov~A., Kuzmenko~E.}
        Texts in, meaning out: neural language models in semantic similarity task for Russian. 
        \textit{arXiv preprint arXiv:1504.08183}. 2015. URL: \url{https://arxiv.org/abs/1504.08183} (access date: 3.04.2018). 

\bibitem{Lesot2008similarity}
    \textit{Lesot~M-J., Rifqi~M., Benhadda~H.} Similarity measures for binary and numerical data: a survey. \textit{International Journal of Knowledge Engineering and Soft Data Paradigms}. 2009. Vol.~1. no.~1. P.~63--84. doi: 10.1504/ijkesdp.2009.021985. URL: \url{http://citeseerx.ist.psu.edu/viewdoc/download?doi=10.1.1.212.6533&rep=rep1&type=pdf} (access date: 3.04.2018). 

\bibitem{Nielsen2018}
\textit{Nielsen~F.}
        Linking ImageNet WordNet Synsets with Wikidata. \textit{In WWW ’18 Companion: The 2018 Web Conference Companion}. 2018. URL: \url{https://arxiv.org/pdf/1803.04349.pdf} (access date: 18.04.2018). 



\end{thebibliographyen}

\vspace*{-0.1cm}
 \begin{flushright}
{\sl\small Received March 31, 2018}
 \end{flushright}
\bfullwidth
\ \vspace*{-0.3cm}
\efullwidth


\end{articletext}

\section{СВЕДЕНИЯ ОБ АВТОРАХ: \hspace{56pt}  CONTRIBUTORS:}
\begin{aboutauthors}
\authorsname{Кириллов Александр Николаевич}
ведущий научный сотрудник, д. ф.-м. н.\\
Институт прикладных математических исследований КарНЦ РАН, Федеральный исследовательский центр <<Карельский научный центр РАН>>\\
ул. Пушкинская, 11, Петрозаводск, \\ 
Республика Карелия, Россия, 185910 \\
эл. почта: kirillov@krc.karelia.ru\\
тел.: (8142) 766312

\columnbreak

\authorsname{Kirillov, Alexander}
Institute of Applied Mathematical Research,\\ Karelian Research Centre,\\ Russian Academy of Sciences\\
11 Pushkinskaya St., 185910 Petrozavodsk,\\ Karelia, Russia\\
e-mail: kirillov@krc.karelia.ru\\
tel.: (8142) 766312
\end{aboutauthors}

\begin{aboutauthors}
\authorsname{Крижановская Наталья Борисовна}
ведущий инженер-исследователь\\
Институт прикладных математических исследований КарНЦ РАН, Федеральный исследовательский центр <<Карельский научный центр РАН>>\\
ул. Пушкинская, 11, Петрозаводск, \\ 
Республика Карелия, Россия, 185910 \\
эл. почта: nataly@krc.karelia.ru\\
тел.: (8142) 766312

\columnbreak

\authorsname{Krizhanovskaya, Natalia}
Institute of Applied Mathematical Research,\\ Karelian Research Centre,\\ Russian Academy of Sciences\\
11 Pushkinskaya St., 185910 Petrozavodsk,\\ Karelia, Russia\\
e-mail: nataly@krc.karelia.ru\\
tel.: (8142) 766312
\end{aboutauthors}

\begin{aboutauthors}
\authorsname{Крижановский Андрей Анатольевич}
рук. лаб. информационных компьютерных \\
        технологий, к. т. н.\\ 
Институт прикладных математических исследований КарНЦ РАН, Федеральный исследовательский центр <<Карельский научный центр РАН>>\\
ул. Пушкинская, 11, Петрозаводск, \\
Республика Карелия, Россия, 185910 \\
эл. почта: andew.krizhanovsky@gmail.com\\
тел.: (8142) 766312

\columnbreak

\authorsname{Krizhanovsky, Andrew}
Institute of Applied Mathematical Research,\\ Karelian Research Centre,\\ Russian Academy of Sciences\\
11 Pushkinskaya St., 185910 Petrozavodsk,\\ Karelia, Russia\\
e-mail: andew.krizhanovsky@gmail.com\\
tel.: (8142) 766312
\end{aboutauthors}

\end{document}